\documentclass{article}
\usepackage{graphicx}
\usepackage[small]{subfigure,epsfig}
\usepackage {amsmath} \usepackage{amssymb}
 \usepackage{cite}
\usepackage{geometry} %flexible way to change page orientation and margins
\begin{document}
%\begin{frontmatter}

\title{Exact solutions of the Kudryashov--Sinelshchikov equation}
\author{Pavel N. Ryabov\footnote{pnryabov@mephi.ru}}

\date{Department of Applied Mathematics, National  Research Nuclear University MEPHI, 31 Kashirskoe
Shosse, 115409 Moscow, Russian Federation}

\maketitle

\begin{abstract}
The Kudryashov-Sinelshchikov equation for describing the pressure waves in liquid with gas bubbles is studied. New exact solution of this equation are found. Modification of truncated expansion method is used for obtaining exact solution of this equation.
\end{abstract}

\emph{Keywords:} Nonlinear evolution equation; Kudryashov-Sinelshchikov equation; Ordinary differential equation; Exact solution.

PACS 02.30.Jr - Ordinary differential equations
\\
%\end{frontmatter}

%-----------------------------------------ÏÎÑÌÎÒÐÅÒÜ----------------------------------------

\section{Introduction}
Recently, Kudryashov and Sinelshchikov \cite{KudryashovSinelshchikov2010} introduced the following equation
\begin{equation}
u_{t}+\gamma u\,u_{x}+\,u_{xxx}-  \varepsilon \,(u\,u_{xx})_{x}-\kappa \,u_{x} u_{xx}-\nu \,u_{xx}-\delta (u\,u_{x})_{x} =0,
\label{KS_1}
\end{equation}
where $\gamma, \varepsilon, \kappa, \nu$ and $\delta$ are real parameters.
Eq. \eqref{KS_1} describes pressure waves in the liquid with gas bubbles taking into account the heat transfer and viscosity \cite{KudryashovSinelshchikov2010}. We call this equation the Kudryashov-Sinelshchikov equation.

It is well known that pressure waves in gas-liquid mixture is characterized by the Burgers-Korteweg-de Vries (BKdV) equation and the Korteweg-de Vries (KdV) equation \cite{Korteweg, Whitham, Nakoryakov, Kudryashov2005}. The Kudryashov-Sinelshchikov equation is generalization of the KdV and the BKdV equation. Indeed, assuming $\varepsilon=\kappa=\delta=0$ we have the Burgers-Korteweg-de-Vries equation. In the case of $\varepsilon=\kappa=\lambda=\delta=0$ we get the famous Korteweg-de Vries equation.

The aim of this work is to find exact solutions of the Kudryashov-Sinelshchikov equation.

We know some methods for finding exact solution of ordinary partial differential(ODE) equations. Let us note some of them: the truncated expansion method \cite{Weiss83,Kudryashov88,Weiss83B, Kudryashov90}, the simplest equation method \cite{Kudryashov05,Kudryashov08}, an automated tanh-function method \cite{Parkes}, the polygons method \cite{Kudryashov07} and the Clarkson-Kruskal direct method \cite{ClarksonKruskal}.

For finding exact solution of the Kudryashov-Sinelshchikov equation we use the modification of  truncated expansion method that was introduced in \cite{Kudryashovbook}. Using truncated expansion method for finding exact solutions of ODEs we obtain the overdetermined system of differential equations. As a rule this systems is difficult to solve. The modification of this method allows us to transform this system of differential equations to the system of algebraic equations. As a result we have essential simplification of solutions construction procedure.

This paper organized as follows. In the Section 2 we introduce the method applied. Solitary waves solutions of Eq. \eqref{KS_1} in the partial case of $\nu=\delta=0$ are described in Section 3. Exact solution of Eq. \eqref{KS_1} in general case are discussed in Section 4.

\section{Method applied}
Let us present the modification of the truncated expansion method \cite{Kudryashovbook}.
We consider the nonlinear partial differential equation in the form
\begin{equation}
E[u_{t},u_{x},\ldots,x,t]=0.
\label{eq1.1}
\end{equation}
Using traveling wave
\begin{equation}
u(x,t)=y(z), \ z=kx-wt.
\label{eq: tr_wave}
\end{equation}
from Eq. \eqref{eq1.1} we obtain the ordinary nonlinear differential equation
\begin{equation}
L[y,y_{z},\dots,k,w]=0.
\label{eq1.3}
\end{equation}

The  modification of the truncated expansion method contains from the following steps \cite{Kudryashovbook}.

\emph{The first step}. Determination of the dominant term with highest order of singularity.
To find dominant terms we substitute
\begin{equation}
y=z^{-p},
\end{equation}
into all terms of Eq. \eqref{eq1.3}. Then we should compare degrees of all terms of Eq. \eqref{eq1.3} and choose two or more with the highest degree. The maximum value of $p$ is called the pole of Eq. \eqref{eq1.3} and we denote it as $N$. It should be noted that method can be applied when $N$ is integer. If the value $N$ is noninteger one can transform the equation studied.

\emph{The second step}. We look for exact solution of Eq. \eqref{eq1.3} in the form
\begin{equation}
y=a_0+a_1Q(z)+a_2(z)Q(z)^2+...+a_{N}Q(z)^{N},
\label{eq2.5}
\end{equation}
where $Q(z)$ is the following function
\begin{equation}
Q(z)=\frac{1}{1+e^z}.
\label{eq2.6}
\end{equation}

\emph{The third step}. We can calculate  necessary number of derivatives of function $y$. It is easy to do using Maple or Mathematica package. Using case $N=2$ we have some derivatives of function $y(z)$ in the form

%As example we consider following case: \\
%\emph{In the case $N=2$ we have the derivatives of function $y(z)$ in the form}

\begin{equation}
\begin{gathered}
y=a_0+a_1Q+a_2Q^2,\hfill\\
y_{z}=-a_1Q+(a_1-2a_2)Q^2+2a_2Q^3,\hfill\\
y_{zz}=a_1Q+(4a_2-3a_1)Q^2+(2a_1-10a_2)Q^3+6a_2Q^4.\hfill
%y=a_0+a_1Q+a_2Q^2+a_3Q^3+a_4Q^4,\hfill\\
%y_{z}=-a_1Q+(a_1-2a_2)Q^2+(2a_2-3a_3)Q^3+(3a_3-4a_4)Q^4+4a_4Q^5,\hfill\\
%y_{zz}=a_1Q+(4a_2-3a_1)Q^2+(2a_1-10a_2+9a_3)Q^3+(6a_2-21a_3+16a_4)Q^4+\hfill\\
%+(12a_3-36a_4)Q^5+20a_4Q^6.\hfill
%y_{zzz}=-a_1Q+(7a_1-8a_2)+(38a_2-27a_3-12a_1)Q^3+\left(6a_1-54a_2+111a_3-\right.\hfill\\
%\left.-64a_4\right)Q^4+(24a_2-144a_3+244a_4)Q^5+(60a_3-300a_4)Q^6+120a_4Q^7,\hfill\\
%y_{zzzz}=a_1Q+(16a_2-15a_1)Q^2+(50a_1-130a_2+81a_3)Q^3+\left(330a_2-60a_1-\right.\hfill\\
%\left.-525a_3+256a_4\right)Q^4+(24a_1-336a_2+1164a_3-1476a_4)Q^5+\left(120a_2-\right.\hfill\\
%\left.-1080a_3+3020a_4\right)Q^6+(360a_3-2640a_4)Q^7+840a_4Q^8.\hfill
\end{gathered}
\label{eq2.7}
\end{equation}

\emph{The fourth step}. We substitute expressions  \eqref{eq2.5}-\eqref{eq2.7} in Eq. \eqref{eq1.1}. Then collect all terms with the same powers of function $Q(z)$ and equate this expressions to zero. As a result we obtain algebraic system of equations. Solving this system we get the values of unknown parameters.

This algorithm can be easily generalized to polynomial differential equation of any order.

\section{Exact solutions of the Kudryashov-Sinelshchikov equation in the case of $\nu=\delta=0$}

Let us find the exact solutions of the Kudryashov-Sinelshchikov equation. Using scale transformation
\begin{equation}
x=x', \quad t=t', \quad u=\frac{1}{\varepsilon}u',
\label{trasformation}
\end{equation}
the Kudryashov-Sinelshchikov equation is written in the form
\cite{KudryashovSinelshchikov2010}
\begin{equation}
u_{t}+\alpha u\,u_{x}+\,u_{xxx}-  \,(u\,u_{xx})_{x}-\beta\,u_{x} u_{xx} =0,
\label{eq: third_or}
\end{equation}
where $\alpha=\gamma/\varepsilon$, $\beta=\kappa/\varepsilon$ (primes are omitted).
Taking the traveling wave ansatz \eqref{eq: tr_wave} into account and integrating with respect to $z$ from Eq. \eqref{eq: third_or} we have
\begin{equation}
C_{1}-\omega\,y+\frac{k\,\alpha}{2}\,y^{2}+k^{3}(y_{zz}-y\,y_{zz}-\frac{\beta}{2}\,y_{z}^{2})=0
\label{r_th_or}
\end{equation}
Here $C_{1}$ is integration constant.

The pole order of Eq. \eqref{r_th_or} is $N=-\frac{2}{\beta+2}$. One can see that at $\beta=-3$ or $\beta=-4$ Eq. \eqref{r_th_or} have the pole of second or first order consequently. So we look for solution of Eq. \eqref{r_th_or} in the form
\begin{equation}
y=a_{0}+a_{1}\,Q+a_{2}\,Q^{2}
\label{r_th_or_exp}
\end{equation}

Substituting \eqref{r_th_or_exp} into Eq.  \eqref{r_th_or} and taking into account relations \eqref{eq2.7} we obtain the system of algebraic equations in the form
\begin{equation}
\begin{gathered}
-2\,k^{3}a_{2}^{2} \left( \beta+3 \right)=0,\vspace{0.2cm} \\
-2\,k^{3}a_{2}\, \left( -2\,a_{2}\,\beta-5\,a_{2}+4\,a_{1}+\beta\,a_{1} \right)=0,\vspace{0.2cm} \\
-\frac{k}{2} \left( 8\,{a_{2}}^{2}{k}^{2}+4\,{k}^{2}{a_{1}}^{2}-26\,{
k}^{2}a_{2}\,a_{1}-12\,a_{2}\,{k}^{2}+12\,{k}^{2}a_{2}\, a_{0}+\right.\\ \left.+4\,{k}^{2}\beta\,a_{2}^{2}+{k}^{2}\beta\,a_{1}^{2}-8
\,{k}^{2}\beta\,a_{2}\,a_{1}-\alpha\,{a_{2}}^{2} \right)=0, \vspace{0.2cm} \\
-k \left( 5\,{k}^{2}a_{2}\,a_{1}+2\,{k}^{2}a_{1}\,a_{0}+2
\,{k}^{2}\beta\,a_{2}\,a_{1}-10\,{k}^{2}a_{2}\,a_{0}-2\,{k
}^{2}a_{1}-\right.\\ \left.-{k}^{2}\beta\,a_{1}^{2}-3\,{k}^{2}a_{1}^{2}+10
\,a_{2}\,{k}^{2}-\alpha\,a_{1}\,a_{2} \right)=0,\vspace{0.2cm} \\
\frac{k\alpha\,a_{1}^{2}}{2}-\omega\,a_{2}-\frac{k^{3}\beta\,
a_{1}^{2}}{2}-3\,k^{3}a_{1}-4\,k^{3}a_{2}\,a_{0}+k\alpha\,
a_{0}\,a_{2}+\\+3\,k^{3}a_{1}\,a_{0}+4\,k^{3}a_{2}-k
^{3}a_{1}^{2}=0,\vspace{0.2cm} \\
-a_{1} \left( \omega-a_{0}\,\alpha\,k+a_{0}\,{k}^{3}-{k}^
{3} \right)=0, \vspace{0.2cm} \\
C_{1}-\omega\,a_{0}+\frac{k\alpha\,a_{0}^{2}}{2}=0
\label{th_or_system}
\end{gathered}
\end{equation}

From \eqref{th_or_system} we have following values of coefficients $a_{0},a_{1},a_{2}$ and paraments $\alpha,\beta,\omega,C_{1}$
\begin{equation}
\begin{gathered}
\beta=-3, \quad  C_{1}=\frac{ \left( {k}^{3}-\omega \right)  \left( 2\,\omega\,{
k}^{2}-\omega\,\alpha-{k}^{3}\alpha \right) }{2\,k \left( {k}^{2}-
\alpha \right) ^{2}},\vspace{0.1cm} \\
a_{2}=\frac {12 k \left( k\,\alpha-\omega \right) }{k^{4}-\alpha^{2}}, \quad
a_{1}=\frac {12 k \left( \omega-k\,\alpha \right) }{{k}^{4}-{\alpha}^{2}}, \quad
a_{0}=\frac {k^{3}-\omega}{k \left( {k}^{2}-\alpha \right) }, \\
\alpha \neq \pm k^{2}
\label{coeffs_th_or_1}
\end{gathered}
\end{equation}

\begin{equation}
\begin{gathered}
\beta=-3, \quad a_{2}=-a_{1} \quad \alpha=k^{2}, \quad a_{0}=-\frac{a_{1}}{6}+1,\\
\omega=k^{3}, \quad C_{1}=-\frac{k^{3}(a_{1}^{2}-36)}{72}
\label{coeffs_th_or_2}
\end{gathered}
\end{equation}

\begin{equation}
\begin{gathered}
\beta=-3, \quad a_{2}=-a_{1} \quad \alpha=-k^{2}, \quad a_{0}=1,\\
\omega=-k^{3}, \quad C_{1}=-\frac{k^{3}}{2}
\label{coeffs_th_or_3}
\end{gathered}
\end{equation}

\begin{equation}
\begin{gathered}
a_{2}=0, \quad \beta=-4, \quad a_{1}=2\,(1-a_{0}), \quad a_{0}\neq 1 \vspace{0.1cm} \\
\alpha=k^{2}, \quad \omega=k^{3}, \quad C_{1}=\frac{a_{0}\,k^{3}(2-a_{0})}{2}
\label{coeffs_th_or_4}
\end{gathered}
\end{equation}

We have four families of solitary wave solutions of Eq. \eqref{r_th_or} in the form
\begin{equation}
y=\frac {k^{3}-\omega}{k \left( {k}^{2}-\alpha \right) }+\frac {12 k \left( \omega-k\,\alpha \right) }{({k}^{4}-{\alpha}^{2})(1+e^{z})}+\frac {12 k \left( k\,\alpha-\omega \right) }{(k^{4}-\alpha^{2})(1+e^{z})^{2}}
\label{D1}
\end{equation}
\begin{equation}
y=1-\frac{a_{1}}{6}+\frac{a_{1}\,e^{z}}{(1+e^{z})^2}
\label{D2}
\end{equation}
\begin{equation}
y=1+\frac{a_{1}\,e^{z}}{(1+e^{z})^2}
\end{equation}
\begin{equation}
y=a_{0}+\frac{2(1-a_{0})}{1+e^{z}}
\end{equation}
corresponding to values of paraments $\alpha,\beta,\omega,C_{1}$ defined by
relations \eqref{coeffs_th_or_1}, \eqref{coeffs_th_or_2}, \eqref{coeffs_th_or_3} and \eqref{coeffs_th_or_4}.
\begin{figure}[!htb]
\center
\includegraphics[width=6cm]{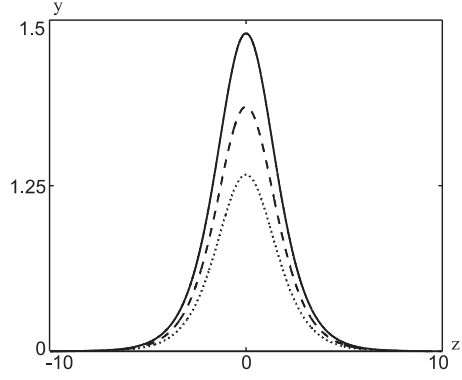}
\caption{{\fontshape{sl}Exact solution of \eqref{D1}, solid -- $k=1, \,\omega=0.75, \,\alpha=2$, dash -- $k=1.1, \,\omega=0.75, \,\alpha=0.5$, dot -- $k=1, \,\omega=1, \,\alpha=0.5$.\selectfont }}\label{p1}
\end{figure}

\begin{figure}[!htb]
\center
\includegraphics[width=6cm]{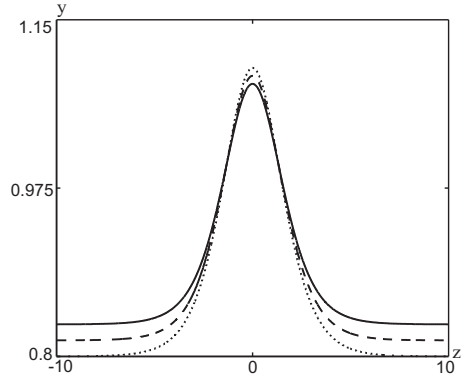}
\caption{{\fontshape{sl}Exact solution of \eqref{D2}, $a_1=1.5; 1; 0.5$ (solid; dash; dot).\selectfont }}\label{p2}
\end{figure}

\section{Exact solutions of the Kudryashov-Sinelshchikov equation in the case of $\nu\neq0, \delta \neq 0$}

Using transformation \eqref{trasformation} we can write the Kudryashov-Sinelshchikov equation in the form
\begin{equation}
u_{t}+\alpha u\,u_{x}+\,u_{xxx}-\,(u\,u_{xx})_{x}-\beta\,u_{x} u_{xx}-\nu\,u_{xx}-\mu (u\,u_{x})_{x} =0,
\label{eq: ext_third_or}
\end{equation}
where $\alpha=\gamma/\varepsilon$, $\beta=\kappa/\varepsilon$, $\mu=\delta/\varepsilon$ (primes are omitted).
Taking traveling wave transformation \eqref{eq: tr_wave} into account and integrating with respect to $z$ from Eq. \eqref{eq: ext_third_or} we have
\begin{equation}
C_{1}-\omega\,y+\frac{k\,\alpha}{2}\,y^{2}+k^{3}(y_{zz}-y\,y_{zz}-\frac{\beta}{2}\,y_{z}^{2})-k^{2}(\nu\,y_{z}+\mu\,y\,y_{z})=0
\label{r_ext_th_or}
\end{equation}

Let us look for solution of Eq. \eqref{r_ext_th_or} in the form
\begin{equation}
y=a_{0}+a_{1}\,Q+a_{2}\,Q^{2}
\label{r_ext_th_or_exp}
\end{equation}

Substituting \eqref{r_ext_th_or_exp} into Eq. \eqref{r_ext_th_or} and with help of relations \eqref{eq2.7} we obtain system of algebraic equations from which we find values of $a_{2},a_{1},a_{0},\omega,C_{1},\beta$
\begin{equation}
\begin{gathered}
\beta=-3, \quad \mu=k ,  \quad a_{2}=-\frac{2(k+\nu)\,k}{\alpha}, \quad a_{1}=\frac{4(k+\nu)\,k}{\alpha}, \\ a_{0}=-\frac {12\,{k}^{2}\nu+12\,{k}^{3}-5\,\alpha\,k+\nu\,\alpha}{6
\alpha\,k}
, \quad
\omega=\frac{\alpha}{6}(5\,k-\nu)-k^{2}(k+\nu),\vspace{0.1cm}\\
C_{1}=\frac { \left( \nu -5\,k \right)  \left( 12\,{k}^{2
}\nu+12\,{k}^{3}-5\,\alpha\,k+\nu\,\alpha \right) }{72k}
\label{r_ext_th_or_coeffs_1}
\end{gathered}
\end{equation}

\begin{equation}
\begin{gathered}
\beta=-3, \quad \alpha=\mu^{2}-k^{2}  , \quad a_{2}=\frac{k^{2}(\nu+\mu)}{\mu(k^{2}-\mu^{2})}, \quad a_{1}=-\frac{(\nu+\mu)\,k}{(k-\mu)\mu}, \\ a_{0}=\frac {2k-\mu+\nu}{2(k-\mu)}, \quad
\omega=-\frac{k}{2}\, \left( 2\,{k}^{2}+\nu\,\mu-{\mu}^{2} \right),\vspace{0.1cm}\\
C_{1}=\frac{k}{8}\left((\mu-\nu)^{2}-4\,k^{2}\right)
\label{r_ext_th_or_coeffs_2}
\end{gathered}
\end{equation}

\begin{equation}
\begin{gathered}
\beta=-4, \quad a_{2}=0, \quad a_{1}=\frac{2(\mu+\nu)\,k}{\mu^{2}-k^{2}+\alpha}, \vspace{0.1cm}\\
 a_{0}=\frac {{k}^{2}+\mu\,k-\alpha+k\nu+\nu\,\mu}{{k}^{2}-\alpha-{\mu}^{2}},\quad
\omega=\frac {k\alpha\, \left( {k}^{2}-\alpha+\nu\,\mu \right) }{{k}^{2}-
\alpha-{\mu}^{2}},\vspace{0.1cm}\\
C_{1}=\frac {k\alpha\, \left({\nu}^{2}{\mu}^{2} -2\,\alpha\,\nu\,\mu-
k^{2}(\mu^{2}+\nu^{2})+ (k^{2}-\alpha)^{2} \right) }{ 2\left( {k}^{2}-\alpha-{\mu}^{2} \right) ^{2}}
\label{r_ext_th_or_coeffs_3}
\end{gathered}
\end{equation}

Using values of parameters \eqref{r_ext_th_or_coeffs_1} we have following kink-type solution of Eq. \eqref{r_ext_th_or_exp}
\begin{equation}
y=-\frac {12\,{k}^{2}\nu-5\,k\alpha+12\,{k}^{3}+\nu\,\alpha}{6\,k
\alpha}+\,\frac {4 \left( k+\nu \right) k}{\alpha\, \left( 1+{
{\rm e}^{z}} \right) }-\,\frac {2 \left( k+\nu \right) k}{\alpha\,
 \left( 1+{{\rm e}^{z}} \right) ^{2}}
\label{r_ext_th_or_sl_1}
\end{equation}

With help of relations \eqref{r_ext_th_or_coeffs_2} we obtain following solution of Eq. \eqref{r_ext_th_or_exp}
\begin{equation}
y=\frac {2\,k-\mu+\nu}{2(k-\mu)}-{\frac { \left( \mu+\nu \right) k}
{ \left( k-\mu \right) \mu\, \left( 1+{{\rm e}^{z}} \right) }}+{
\frac {{k}^{2} \left( \mu+\nu \right) }{\mu\, \left( {k}^{2}-{\mu}^{2}
 \right)  \left( 1+{{\rm e}^{z}} \right) ^{2}}}
\label{r_ext_th_or_sl_2}
\end{equation}
Taking into account values of parameters \eqref{r_ext_th_or_coeffs_3} we have following kink-type solution of Eq. \eqref{r_ext_th_or_exp}
\begin{equation}
y=\frac {{k}^{2}+\mu\,k-\alpha+k\nu+\nu\,\mu}{{k}^{2}-\alpha-{\mu}^{2}}+\frac{2(\mu+\nu)\,k}{(\mu^{2}-k^{2}+\alpha)(1+e^{z})}
\label{r_ext_th_or_sl_3}
\end{equation}

\begin{figure}[!htb] %[!ht]
\centering         %width=8.3cm, height=5.6cm
\includegraphics[width=6cm]{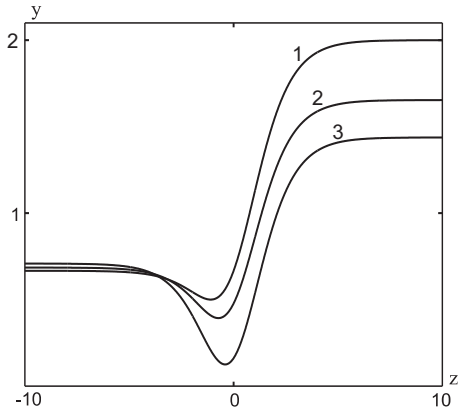}
\caption{The solution \eqref{r_ext_th_or_sl_2} of Eq. \eqref{r_ext_th_or_exp} at $\mu=1; 0.7; 0.4$ (curves 1,2,3).}
\label{fig: ext_th_or}
\end{figure}

Dependence solution \eqref{r_ext_th_or_sl_2} from $z$ at different values of parameter $\mu$ at $k=2,\nu=1$ are illustrated on Fig. \ref{fig: ext_th_or}.

We believe that solutions  \eqref{r_ext_th_or_sl_1}, \eqref{r_ext_th_or_sl_2} and \eqref{r_ext_th_or_sl_3} are new.

\section{Conclusion}
The Kudryashov-Sinelshchikov equation was studied using the modification of the truncated method.
The algorithm of the method applied was presented. The efficiency of this method was demonstrated. New exact solution of the Kudryashov-Sinelshchikov equation were obtained.

This work was supported by the federal target programm "Research and scientific-pedagogical personnel of innovation in Russia" on 2009-2011, Contracts P 28, P 741.

\end{document}